\documentclass[aps,prl,reprint,groupedaddress]{revtex4-1}

\usepackage{graphicx}
\usepackage{amsmath}

\begin{document}

\title{Experimental realization of a Coulomb blockade refrigerator}

\author{A. V. Feshchenko} \email{anna.feshchenko@aalto.fi}
\affiliation{Low Temperature Laboratory, O.V. Lounasmaa Laboratory, Aalto University, FI-00076  Aalto, Finland}
\author{J. V. Koski}
\affiliation{Low Temperature Laboratory, O.V. Lounasmaa Laboratory, Aalto University, FI-00076  Aalto, Finland}
\author{J. P. Pekola}
\affiliation{Low Temperature Laboratory, O.V. Lounasmaa Laboratory, Aalto University, FI-00076  Aalto, Finland}

\date{\today}

\begin{abstract}
We present an experimental realization of a Coulomb blockade refrigerator (CBR) based on a single - electron transistor (SET). In the present structure, the SET island is interrupted by a superconducting inclusion to permit charge transport while preventing heat flow. At certain values of the bias and gate voltages, the current through the SET cools one of the junctions. The measurements follow theoretical model down to $\sim$  80~mK, which was the base temperature of the current measurements. The observed cooling increases rapidly with decreasing temperature in agreement with the theory, reaching about 15 mK drop at the base temperature. CBR appears as a promising electronic cooler at temperatures well below 100 mK.
\end{abstract}


\maketitle

Over the past several decades, there has been continuous growth of research dedicated to thermoelectrics in a variety of nanostructures, see Refs. \cite{Rowe2006, Mahan1994, Shakouri2006} and references therein, and in particular to electronic microrefrigerators \cite{Muhonen2012, Courtois2014}. The first quantum dot refrigerator (QDR) for cryogenic temperatures was proposed by Edwards in 1993 \cite{Edwards1993, EdwardsReview1995}. In the QDR discrete energy levels of the dot are to be tuned to cool the electronic Fermi-Dirac distribution of a small reservoir. It was soon after demonstrated in a different device \cite{Nahum1994}, that the electrons can be cooled down below the phonon temperature of the lattice. These cooling effects have been observed in several different systems, such as a normal metal-insulator-superconductor (NIS) tunnel junction \cite{Nahum1994, Fisher1999, Clark2004} and with improved performance in a related SINIS device with a double junction configuration \cite{Leivo1996, Saira2007}. Also the exact realization of the above mentioned QDR was shown later in a 2D electron gas \cite{Prance2009}. All these devices which are able to cool down on-chip electronic systems to subkelvin temperatures have potential scientific and commercial applications, especially when size, weight and ease of operation become important. Examples of possible applications are microbolometers discussed in Ref. \cite{Giazotto2006}. Several implemented coolers on a Si$_{3}$N$_{4}$ membrane \cite{Clark2005} with a transition-edge sensor for high-resolution x-ray spectroscopy \cite{Miller2008} have been demonstrated recently. These refrigerators work most of the time in the temperature range of 300 - 100 mK. They perform sub-optimally at lower temperatures, due to several reasons, including excess quasi-particle population in a superconductor \cite{Knowles2012, Klapwijk2011}, leakage of the junctions \cite{Pekola2004} and low electron-electron scattering rate \cite{Prance2009}.

In this Letter, we present the first experimental realization of a new type of a cooler that has been recently proposed by some of the authors \cite{Pekola2014}. A Coulomb blockade refrigerator (CBR) is based on thermal transport through a fully normal single-electron transistor (SET) \cite{Averin1986}. The relative temperature drop is expected to increase when the base temperature is lowered, which we verify here experimentally down to 80 mK. This feature makes the CBR suited for temperatures below 100 mK, and could be used in cascade coolers as the last stage, for example in combination with a superconducting refrigerator \cite{Quaranta2011}.  Here we want to emphasize the most important feature of the CBR. In contrast to the NIS cooler, the operating point can be optimized by the bias and the gate voltages at any given temperature. For example, at a base temperature of 20 mK, which is possible to achieve with standard dilution refrigerators, the temperature of $\sim$ 5 mK can be reached by CBR assuming no extra heat leaks.

\begin{figure}[h!t]
\includegraphics[width=0.45\textwidth]{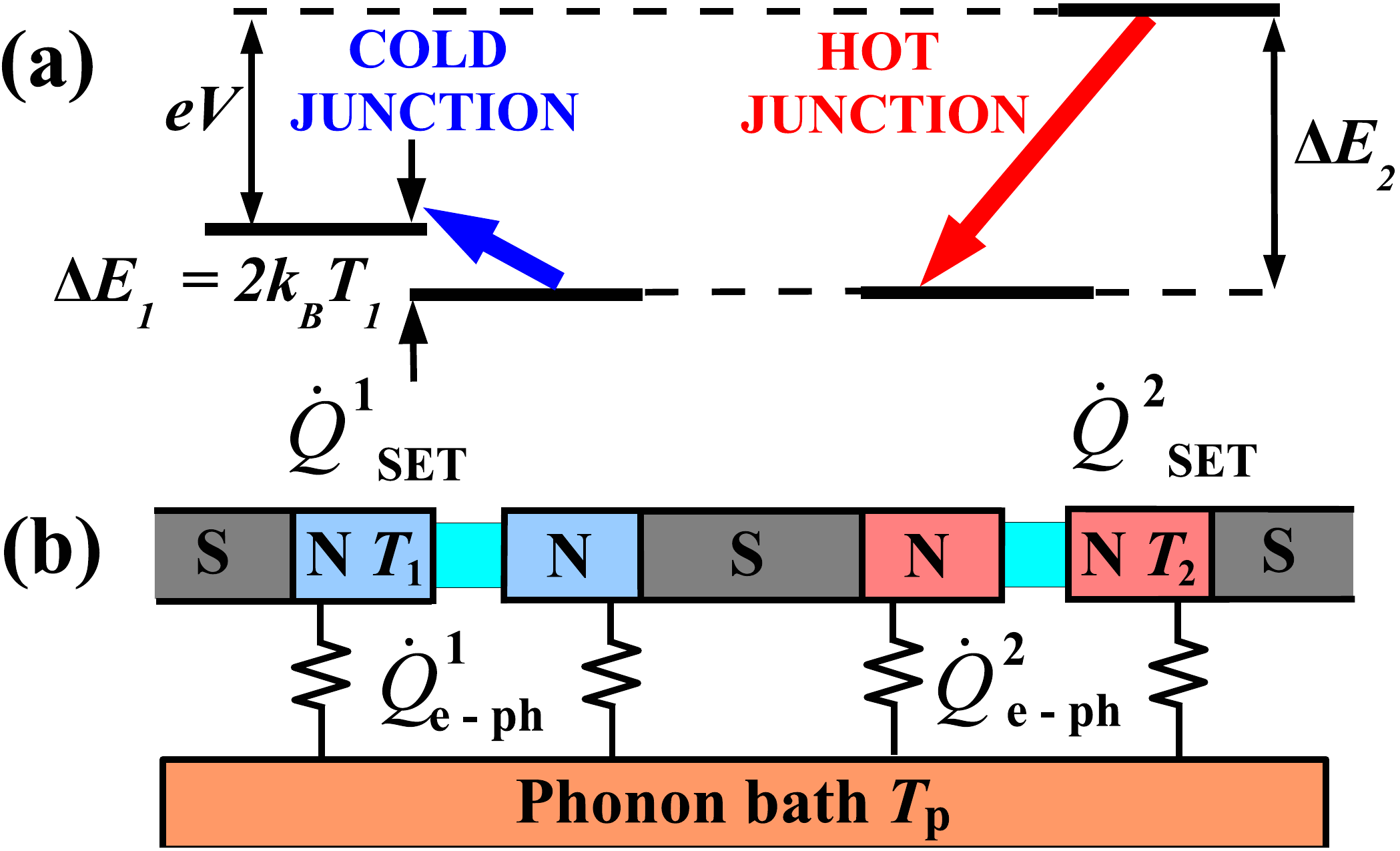}
\caption{(Color on-line) Principle of operation of the cooler. (a) Schematic of the energy levels of the device at its operation point. (b) Thermal scheme of the structure.}
\label{fig:1}
\end{figure}

Figure ~\ref{fig:1} shows the principle of operation of the cooler. The SET has two NIN tunnel junctions with tunneling resistances $R_{T, k}$ each, formed between two normal metal (N) electrodes separated by a thin insulator (I) layer. We define the order of the junctions as $k$ = 1,2, where 1 is the "cold" junction and 2 is the "hot" junction. Our cooler is biased by voltage $V$ transporting electrons through the SET, first to the island through the ''hot'' junction, then out of the island through the ''cold'' junction. When the SET island has $n$ excess electrons, the electrostatic energy is $E = E_{c}(n - n_{g})^{2}$, where  $E_{c} = e^{2}/2C_{\Sigma}$ is the charging energy, and $C_{\Sigma}$ is the total capacitance of the SET island. The electrostatic energy is controlled by tuning the gate position $n_{g} \equiv - C_{g}V_{g}/e$, where $C_{g}$  and $V_{g}$ are the gate capacitance and voltage, respectively. For simplicity, we define the two extreme gate values that will be used later: (i), gate closed ($n_{g}$ = 0) and (ii), gate open position ($n_{g}$ = 0.5).

In this experiment, we focus on low-temperature regime, where the number of excess electrons on the island is restricted to $n$ = 0 or $n$ = 1. An electron that tunnels into the island through junction 2 changes $n$ from 0 to 1 with an energy cost $\Delta E_{2} =  eV/2 + E_{c}(1-2n_{g})$. Similarly an electron that tunnels out of the island through junction 1 changes $n$ from 1 to 0 for an energy cost $\Delta E_{1} = eV/2 - E_{c}(1-2n_{g})$. The tunneling electrons distribute the energy evenly to their respective heat baths formed by the junction electrodes with typically about 10$^9$ free electrons in each. The energy $\Delta E_{2}$ is added to the heat bath of junction 2 at temperature $T_{2}$, heating that junction, while the energy $\Delta E_{1}$ is removed from the junction 1 heat bath at temperature $T_{1}$ cooling that junction.
At the optimum point of $n_{g}$ and $V$, where $\Delta E_{1} = 2k_{B}T_{1}$, the cooling power of the junction 1 is given by \cite{Pekola2014}
\begin{equation} 
\dot{Q}_{opt}\simeq0.31\frac{(k_{B}T_{1})^2}{e^2R_{T, 1}}.
\label{Eq.1} 
\end{equation}

Next we consider the conditions to observe a temperature drop in the CBR. As presented in Fig.~\ref{fig:1}~(b), substrate phonons exchange heat with electrons via electron-phonon coupling $\dot{Q}_{e-ph}^{k} = \Sigma \Omega_{k} (T_{k}^{5} - T_{p}^{5})$, where $\Sigma$ is a constant specific to the electrode material, $\Omega_k$ is the volume of that electrode, $T_{k}$ is the temperature of the electrons, and $T_p$ is the temperature of the phonon bath. Since $\dot{Q}_{e-ph}^{k}$ is proportional to electrode volume, the CBR junction electrodes need to be small, as well as thermally insulated to prevent heat leaks from outside and between each other as indicated in Fig.~\ref{fig:1}~(b) by S parts. Finally one needs temperature probes to measure $T_1$ and $T_2$. To realize the CBR, we make use of several different types of high-quality Al junctions all contacting Cu, compatible with co-fabrication of other metallic structures \cite{Clark2005, Miller2008, Quaranta2011}.

\begin{figure}[h!t]
\includegraphics[width=0.5\textwidth]{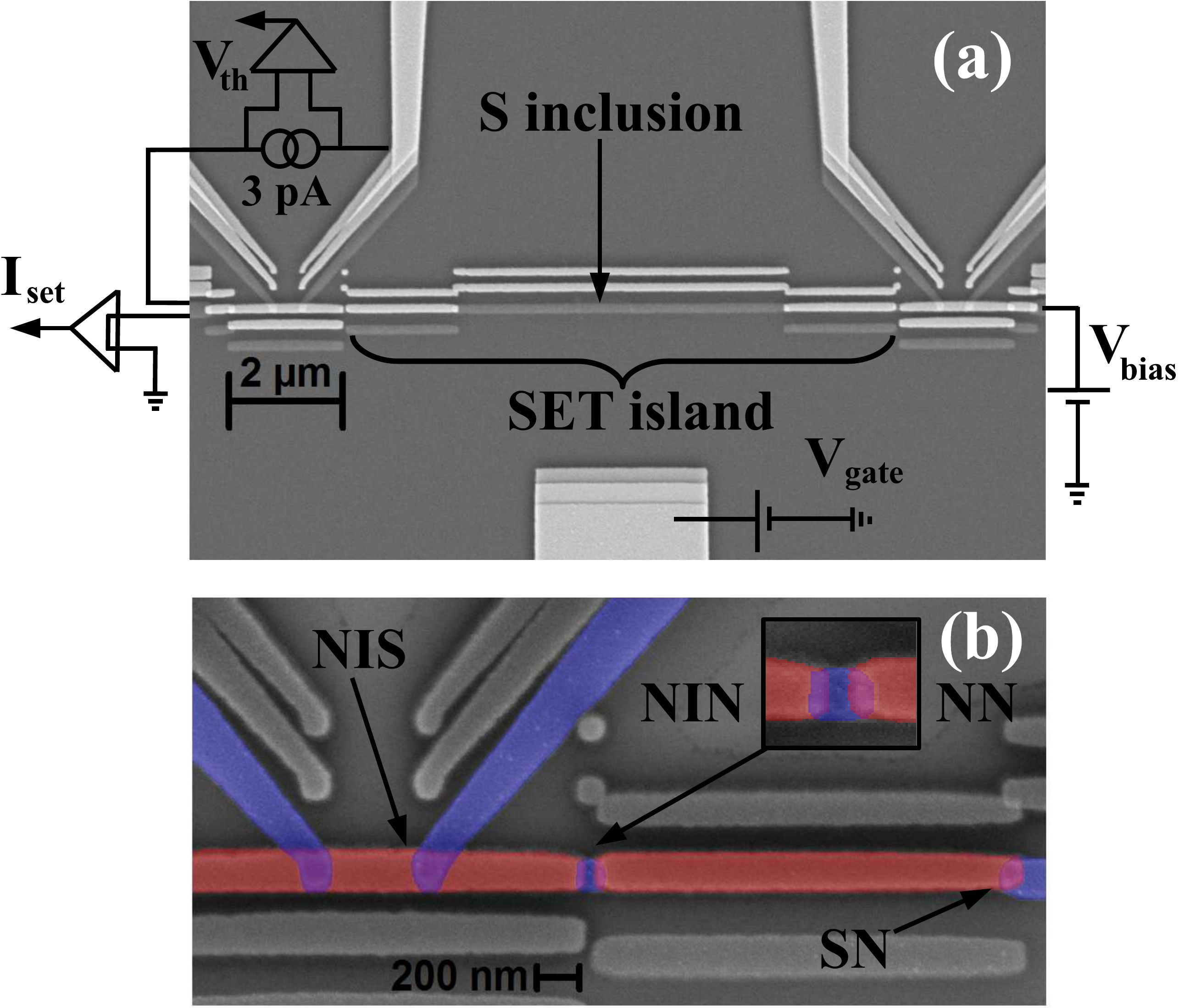}
\caption{(Color on-line) (a) SEM image of the device shown together with a schematic of the experimental setup, where the SET island consists of two normal metal islands with a superconducting inclusion in between. (b) Close-up SEM image shows the left part of the SET that is connected to the cooled island through the Al ''dot'' (see text for more details). Zoom into the Al ''dot'' with NIN and clean NN contacts is shown in the inset. In the main panel, the NIS probes used to monitor the temperature of the cooled island are highlighted. All normal metal parts are colored red, while superconductors are shown in blue.}
\label{fig:2}
\end{figure}

In the following, we describe the present realization of the proof-of-concept CBR that is shown in a scanning-electron micrograph (SEM) in Figs.~\ref{fig:2}~(a) and (b). The device is made by electron beam lithography and three-angle shadow evaporation technique \cite{Fulton1987}. First, a 20 nm layer of Al forms the superconducting inclusion, outer leads, fingers for the NIS temperature probes and Al "dots" [see the inset in Fig.~\ref{fig:2}~(b)]. Next, a 25 nm layer of Cu, evaporated at a different angle, creates normal metal parts that form clean contacts to the superconducting inclusion from one side, and to Al ''dots'' on the other side. Short Cu intermediate leads that are connected to the outer leads are formed at this step as well. Thermal oxidation in-situ followed after the second metal evaporation to form a thin $Al_{2}O_{3}$ layer over Al to provide the insulator for tunnel junctions. In the last evaporation step, 25 nm thick Cu electrodes form tunnel junctions for the SET, for the NIS probes and connections to the intermediate leads. Connections to the outer leads are assumed to be transparent junctions between two Cu layers, which are, however, slightly oxidized in between. Underneath of all the leads, except for the gate electrode, we use a ground plane made out of 50 nm of copper covered by 50 nm of AlO$_{x}$. Here, the ground plane serves the purpose of reducing the leakage current in the sub-gap region of the NIS thermometer \cite{Pekola2010, Saira2012}, and to suppress voltage noise in general.

The SET junctions are made by the laterally proximized tunnel junction technique \cite{Koski2011}, where the small Al ''dots'' (100 nm $\times$ 100 nm) are in direct contact with a bigger volume of normal metal Cu that suppresses the superconductivity by inverse proximity effect \cite{Belzig1996, Belzig1999, Peltonen2010}. The inset in Fig.~\ref{fig:2}~(b), shows the lateral junction with an Al ''dot'' (colored  blue) that is connected on the left side by the NIN tunnel junction to the island to be cooled, and to the SET island through a clean NN contact on the right side. Two fully normal NIN tunnel junctions of the SET were made intentionally un-equal by lithography. Junction 1 with low resistance increases the cooling power at the optimal bias point, see Eq.~(\ref{Eq.1}), and at the same time junction 2 is more resistive decreasing  cotunneling to be described later in the text. The SET island splits into three parts, where the superconducting section of 5 $\mu$m length in the middle is surrounded by shorter normal metal islands that are 2 $\mu$m long each as indicated in Fig.~\ref{fig:2}~(a). We have chosen the particular length of the superconductor for two reasons. First, to avoid the inverse proximity effect, the length of the inclusion is much longer than the coherence length for Al that is typically around 200 nm at 100 mK. Thus the superconductor provides good thermal insulation \cite{Peltonen2010}, while maintaining electrical connection through the SET. Second, to keep the total charging energy of the SET island in the range of our interest, around 1 K, it cannot be longer than a few $\mu$m.

\begin{figure}[h!t]
\includegraphics[width=0.45\textwidth]{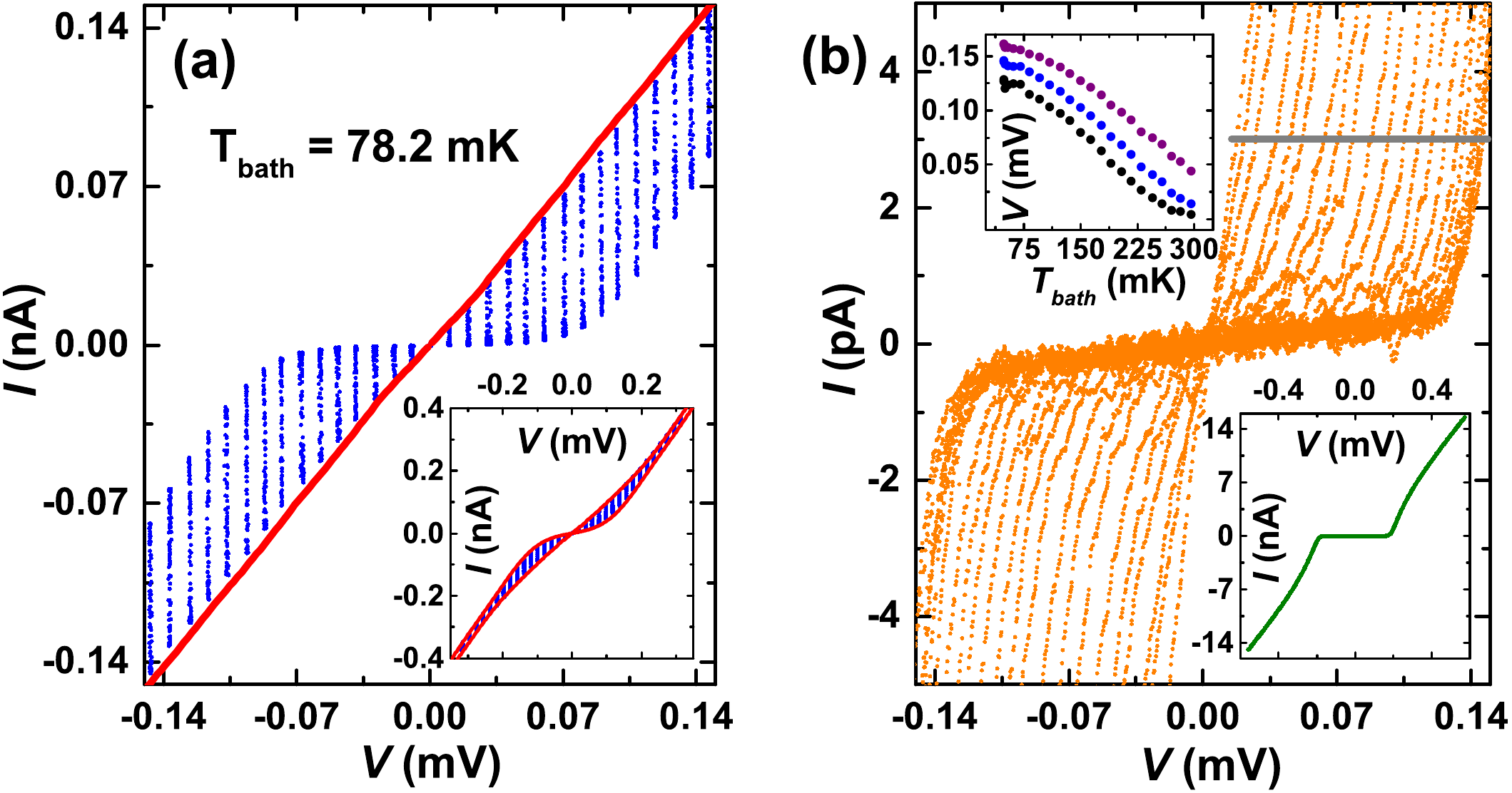}
\caption{(Color on-line) (a) Close-up of the measured  $I$-$V$ of the SET shown in blue dots. The solid red line is the fit for gate open case. The inset shows the full $I$-$V$ of the SET at $\sim$ 207 mK (blue dots) together with a full theory fit (solid red line) for both gate open and gate closed positions utilizing the same $R_{T}$ and $E_{c}$ as in the main panel. (b) $I$-$V$ curves (orange dots) of one of the probing NIS junctions at various temperatures between 50 and 300 mK. The solid horizontal grey line indicates the 3 pA current bias of the NIS probe (see text for details). The lower inset shows the measured full $I$-$V$ at the bath temperature of $\sim$ 48 mK (green dots). The upper inset shows the voltage calibration obtained from the $I-V$ curves in the main panel for bias currents of 1 pA, 3 pA and 14 pA (shown as black, blue and purple dots, respectively).}
\label{fig:3}
\end{figure}

Next, we present the measurement data that characterizes the SET and the NIS temperature probes. We conduct our experiments in a $^3$He - $^4$He dilution refrigerator at a bath temperature $T_{bath}$. In Fig.~\ref{fig:3}~(a), the measured current voltage characteristic ($I$-$V$) of the SET is shown in blue dots. The dots appear as vertical lines due to the gate voltage sweep recorded at each bias voltage point. The slope in the gate open position is constant around zero bias demonstrating the absence of the superconducting gap, meaning that the SET junction electrodes are normal. Theoretical model used in the inset and in the main panel is based on the theory of sequential single-electron tunneling \cite{Averin1986}. The fit to the $I$-$V$ curve gives values $E_{c}$ = 78 $\mu$eV for the charging energy, and the tunneling resistances are $R_{T, 1}$ = 103 k$\Omega$ and $R_{T, 2}$ = 448 k$\Omega$.

In Fig.~\ref{fig:3}~(b), we characterize the NIS thermometer probe by measuring the temperature sweep of its $I$-$V$. The curves are well separated for different temperatures. The solid grey line in Fig.~\ref{fig:3}~(b) indicates 3 pA that we have chosen as a biasing current for the temperature probe. This value is in the temperature sensitive region, not influenced by the Andreev current leakage \cite{Hekking1993, Courtois2008, Greibe2011} that appears as a constant slope at low bias values in Fig.~\ref{fig:3}~(b).

\begin{figure}[h!t]
\includegraphics[width=0.45\textwidth]{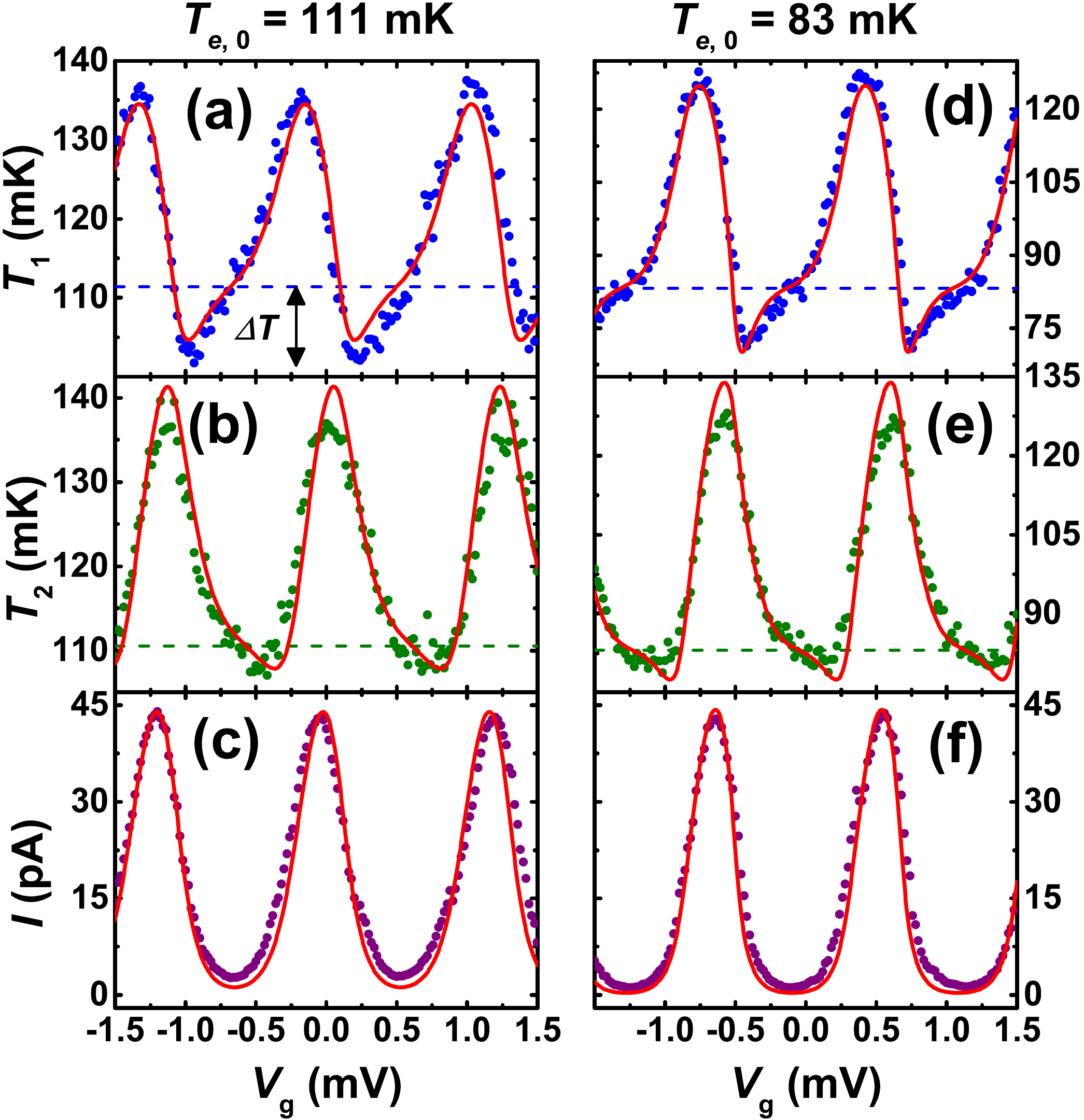}
\caption{(Color on-line) Top panels: temperature $T_{1}$ over few gate periods (shown in blue dots in (a) and (d)). The solid red lines correspond to the theoretical prediction, where the voltage of the SET has a value of 45 $\mu$eV. The base electronic temperatures $T_{e, 0}$ are shown in blue dashed lines for data. Two middle panels, (b) and (e), show $T_{2}$ (green dots) together with the theory (solid red lines), and the base temperatures marked as dashed lines. The two lowest panels (c) and (f), present the gate dependent current $I$ through the SET, measured simultaneously as the temperatures $T_{1}$ and $T_{2}$.}
\label{fig:4}
\end{figure}

\begin{figure}[h!t]
\includegraphics[width=0.45\textwidth]{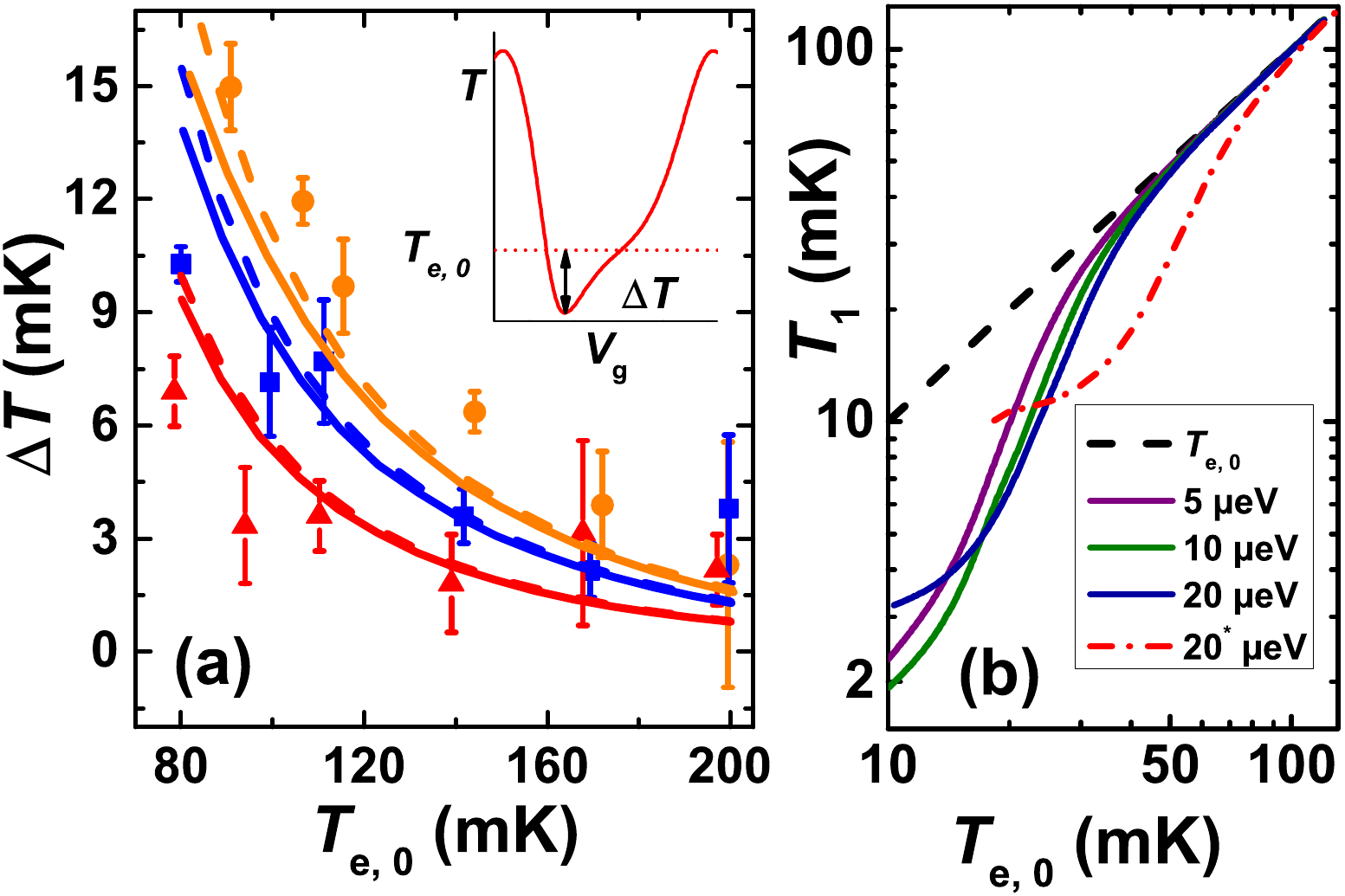}
\caption{(Color on-line) (a) Observed cooling $\Delta T$ (see inset for definition) for three different voltage bias values over the temperature range from 200 mK down to 80 mK. The data sets obtained by averaging over all realizations within $\pm$ 5 $\mu$eV range are shown for the mean bias values of 60 $\mu$eV (orange circles), 40 $\mu$eV (blue squares) and 20 $\mu$eV (red triangles) from top to bottom. 
The error bars are calculated as a standard deviation for the data. The theoretical prediction of $\Delta T$ for the three voltage bias values (60 $\mu$eV, 40 $\mu$eV, 20 $\mu$eV) is shown as orange, blue and red solid lines, respectively. The dashed lines show the theoretical prediction without cotunneling. (b) Theoretical prediction of $T_{1}$ for the three voltage bias values as indicated in the legend in the temperature range of 10...130 mK is shown as solid lines. The dash-dotted red line indicated as 20$^{*}$ $\mu$eV is the same as in panel (a), where in the model we used values of $R_{T, k}$ obtained from the experiment. The solid purple, dark green and dark blue lines are predictions for a CBR with tunneling resistances ten times higher than those in the present measurements.}
\label{fig:5}
\end{figure}

The experimental setup to test CBR performance is shown in Fig.~\ref{fig:2}~(a). The voltage bias is applied to the right lead of the SET and at the same time the gate voltage is used to tune the electrostatic potential of the island. At each bias point, the gate voltage is swept over a few gate periods, at constant current bias for temperature probes. The observed cooling depends on the bath temperature, as well as on the bias and gate voltages. We define the base temperature $T_{e, 0}$ as the electronic temperature observed at the gate closed position meaning zero current through the SET; here no cooling or heating takes place at the junctions. The temperature traces obtained by the NIS thermometer at different bath temperatures are presented in Fig.~\ref{fig:4}. The measured thermal voltages across the NIS probes and the outer leads are converted into temperature with a calibration \cite{Nahum1994, Giazotto2006, Meschke2006, Hung2014} (see upper inset in Fig.~\ref{fig:3}~(b)) with respect to the bath temperature of the dilution refrigerator. The measurement data shown in the left column (panels (a) - (c) in Fig.~\ref{fig:4}) and in the right column (panels (d) - (f)) were obtained at base temperatures of 111 mK and 83 mK, respectively. Both columns from top to bottom present the temperature traces of junction 1 (shown in panels (a) and (d)), and of junction 2 (shown in panels (b) and (e)), and the current through the SET (shown in panels (c) and (f)) during the gate voltage sweep. The cooling $\Delta T$ appears as a dip below $T_{e, 0}$, indicated by the dashed lines, and the signal above the line indicates heating. The solid red lines shown in all the panels present the theoretical model to be described. We measured similar temperature traces at various bath temperatures. At each bath temperature, the drop $\Delta T$ was averaged over repeated gate sweeps and the standard deviation was calculated. The result is shown in Fig.~\ref{fig:5}~(a) for three different voltage bias values indicated by orange circles, blue squares and red triangles. The error bars are $\pm 2\sigma$ confidence intervals estimated from the observed scatter.

Next, we describe the theoretical model presented by the lines in Figs.~\ref{fig:4} and \ref{fig:5}. The electron temperatures $T_{1}$ and $T_{2}$  are obtained as a solution to the steady state heat balance equation $\dot{Q}_{SET, k} + \dot{Q}_{e-ph}^{k} = 0$ for both junctions $k = 1, 2$, where $\dot{Q}_{SET, k}$ is the heat flow through the junctions based on sequential single-electron tunneling model and cotunneling \cite{Pekola2014}. We estimate that in the present structure at $T_{e, 0}$ = 83 mK the temperature drop is reduced by 11 $\%$ due to cotunneling. We assume $T_{p} = T_{e, 0}$ in the expression of electron-phonon coupling $\dot{Q}_{e-ph}^{k}$. The volume of the cooled island is approximately $\Omega \simeq 5 \times 10^{-21}$ m$^{3}$. We obtain a fit $\Sigma = 4 \times 10^{9}$ W K$^{-5}$ m$^{-3}$ by assuming that at gate open position the heat flow through the junctions is equal to Joule heating, $\dot{Q}_{SET, k}$ = $IV/4$, distributed equally among the four electrodes of the SET.

The observed cooling increases rapidly in agreement with the theory towards low temperatures, giving the maximum value of $\Delta T$ = 15 mK $\pm$ 1.15 mK for bias voltage of 60 $\mu$eV at $T_{e, 0}$ = 90 mK. The electronic temperature saturation at 80 mK can be explained by the NIS thermometer losing its sensitivity (see upper inset in Fig.~\ref{fig:3}~(b)). Another contribution is the heating due to radiation from the hotter electromagnetic environment \cite{Hergenrother1995, Pekola2010, Saira2012}. These problems can be resolved in the future by improving filtering and thermometry. Increased resistance of the NIS junctions would reduce the magnitude of Andreev current and the heat it produces which is significant towards low temperatures \cite{Courtois2008}. Alternative thermometry can be used as well. For example, a proximity Josephson junction has temperature dependent critical current \cite{Dubos2001, Courtois2008, Meschke2009} and essentially zero dissipation, and it could be adjusted to a specific temperature interval.

Finally, in Fig.~\ref{fig:5}~(b) we show the theoretical prediction within the model above, including the effect of cotunneling, but with the tunneling resistances that are ten times higher than in the present experiment. As a reference, we show the dash-dotted red line (20$^{*}$ $\mu$eV), which is identical to the one that is shown in panel (a) as red curve, with $R_{T, k}$ values obtained from the experiment. To reach lower electronic temperatures in the present configuration, one can thus benefit from higher tunneling resistances of the junctions and lower bias voltages.

In conclusion, we have shown the experimental realization and demonstrated the proof-of-concept performance of a Coulomb blockade refrigerator. The present realization of the device measured down to 80 mK demonstrates about 15 $\%$ temperature drop which increases rapidly towards lower temperatures.

\begin{acknowledgments}
We acknowledge the availability of the facilities and technical support by Otaniemi research infrastructure for Micro and Nanotechnologies (OMN). We acknowledge financial support from the European Community FP7 Marie Curie Initial Training Networks Action (ITN) Q-NET 264034 and INFERNOS grant (project no. 308850) and the Academy of Finland though its LTQ CoE grant (project no. 250280) and V\"ais\"al\"a Foundation. We thank D. V. Averin and I. M. Khaymovich for useful discussions.
\end{acknowledgments}


%

\end{document}